\documentclass[aps,prl,twocolumn,superscriptaddress]{revtex4-1}

\usepackage{graphicx}
\usepackage{amssymb}
\usepackage{amsmath}
\usepackage{color}
\usepackage{physics}
\usepackage{bbm}
\usepackage{pdfpages}
\usepackage{float}

\renewcommand{\b}{\mathbf}

\makeatletter
\AtBeginDocument{\let\LS@rot\@undefined}
\makeatother

\begin{document}
\title{Floquet Engineering Ultracold Polar Molecules to Simulate Topological Insulators}

\author{Thomas Schuster}
\affiliation{Department of Physics, University of California, Berkeley, California 94720 USA}
\author{Felix Flicker}
\affiliation{Department of Physics, University of California, Berkeley, California 94720 USA}
\affiliation{Rudolph Peierls Centre for Theoretical Physics, University of Oxford, Department of Physics, Clarendon Laboratory, Parks Road, Oxford, OX1 3PU, UK}
\author{Ming Li}
\affiliation{Department of Physics, Temple University, Philadelphia, Pennsylvania 19122, USA}
\author{Svetlana Kotochigova}
\affiliation{Department of Physics, Temple University, Philadelphia, Pennsylvania 19122, USA}
\author{Joel E. Moore}
\affiliation{Department of Physics, University of California, Berkeley, California 94720 USA}
\affiliation{Materials Science Division, Lawrence Berkeley National Laboratory, Berkeley, California 94720, USA}
\author{Jun Ye}
\affiliation{JILA, National Institute of Standards and Technology and Department of Physics,
University of Colorado, Boulder, CO 80309, USA}
\author{Norman Y. Yao}
\affiliation{Department of Physics, University of California, Berkeley, California 94720 USA}
\affiliation{Materials Science Division, Lawrence Berkeley National Laboratory, Berkeley, California 94720, USA}
\date{\today}

\renewcommand{\topfraction}{.85}
\renewcommand{\bottomfraction}{.7}
\renewcommand{\textfraction}{.15}
\renewcommand{\floatpagefraction}{.66}
\renewcommand{\dbltopfraction}{.66}
\renewcommand{\dblfloatpagefraction}{.66}
\setcounter{totalnumber}{1}

\newcommand{\FigureHopf}{
 \begin{figure}[t]
\centering
\includegraphics[width=\columnwidth]{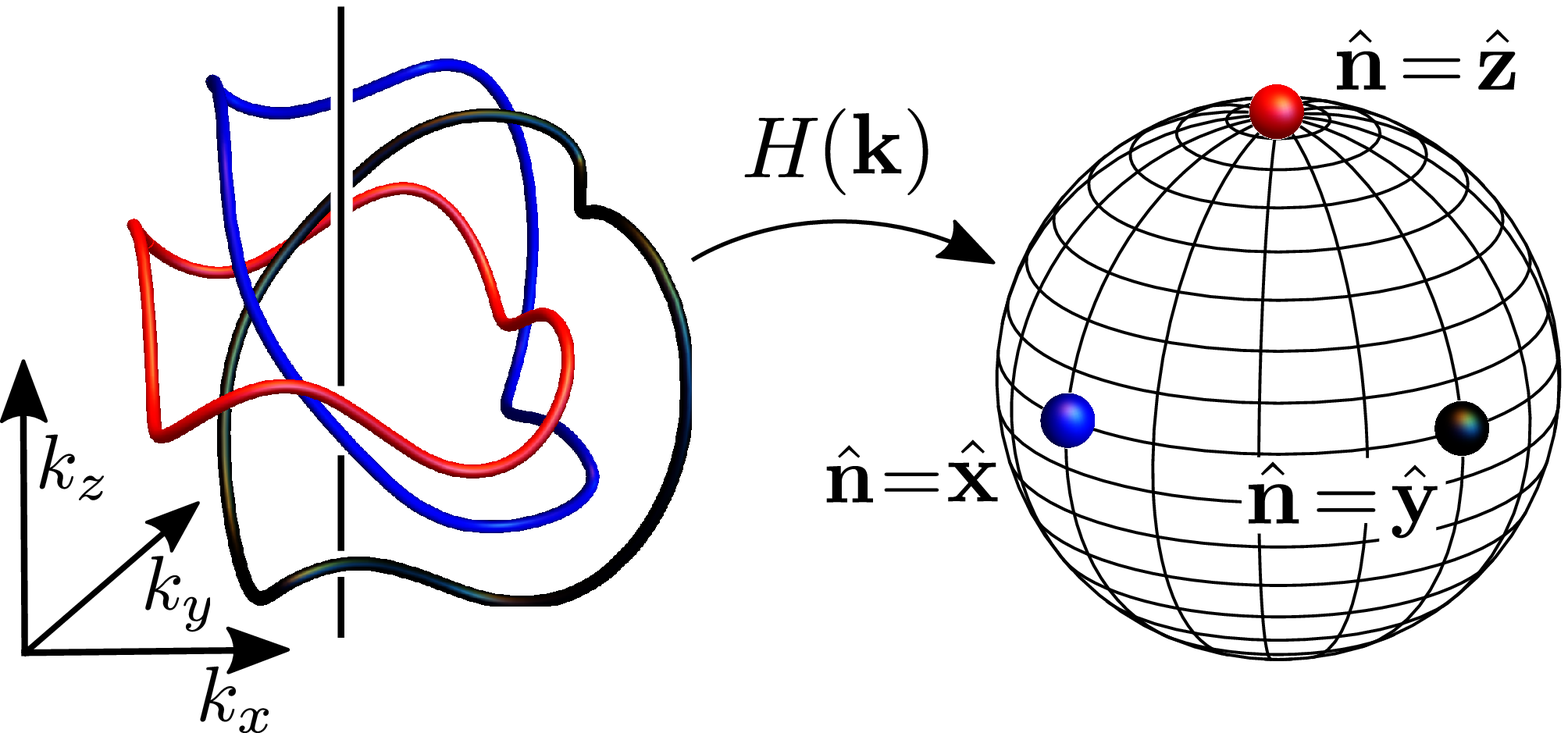}
\caption{
The Hamiltonian of the Hopf insulator maps closed loops in the Brillouin zone to points on the Bloch sphere, with the unique property that any two loops have linking number equal to the Hopf invariant. The above loops are solved for using the proposed experimental dipolar spin Hamiltonian specified in the text -- their linking provides a visual verification of the Hopf insulating phase. The $\hat{\textbf{n}} = \hat{\textbf{x}},\hat{\textbf{y}}$ pre-images (blue, black tubes) are $90^\circ$ rotations of each other about the $k_z$-axis (vertical black line) due to the spin-orbit coupled hopping $t^{AB}_{\b{r}} \sim e^{i \phi}$.
} 
\label{fig: hopf}
\end{figure}
}

\newcommand{\FigureLattice}{
 \begin{figure}[t]
\centering
\includegraphics[width=\columnwidth]{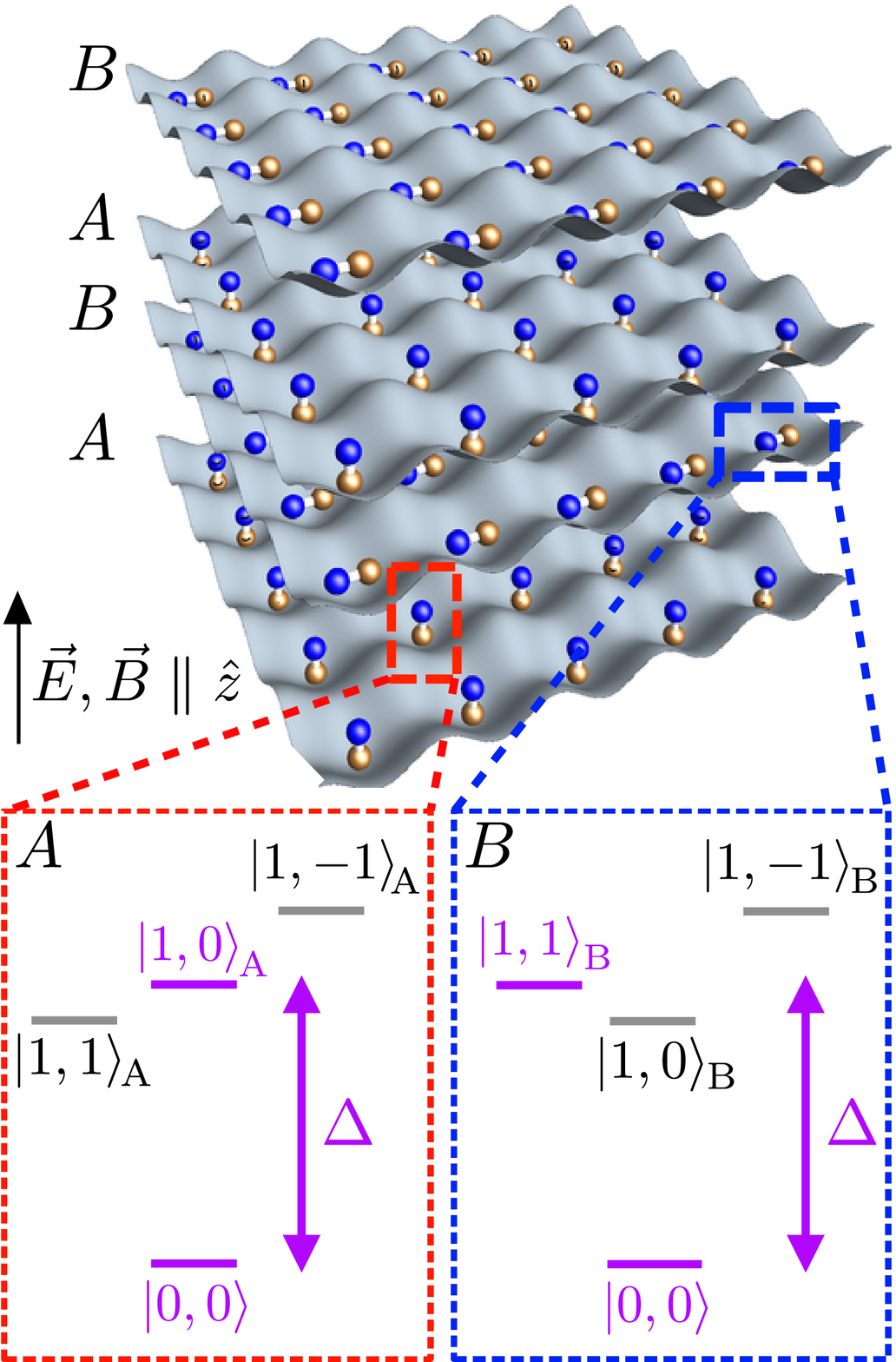}
\caption{
Schematic geometry depicting a 3D optical lattice of polar molecules with two layered sublattices $A$ and $B$. Orbital motion of the molecules is frozen by the optical lattice.
The level structure of the $J=0,1$ rotational states on the $A$ (left) and $B$ (right) sublattices. The purple highlighted states form the hard-core bosonic doublet for each sublattice, and their energy splitting $\Delta$ is tuned by external fields to be degenerate between sublattices.
} 
\label{fig: lattice}
\end{figure}
}

\newcommand{\FigureHopping}{
 \begin{figure}[t]
\centering
\includegraphics[width=\columnwidth]{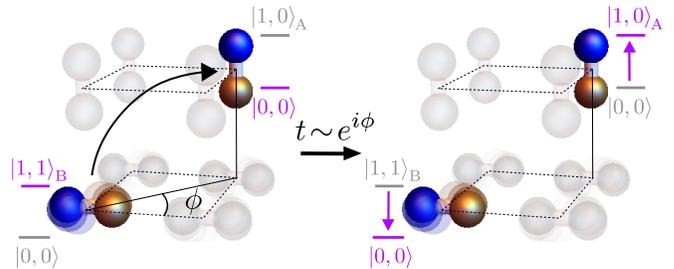}
\caption{
Depiction of the inter-sublattice `hopping' $\ket{0,0}_A \ket{1,1}_B \rightarrow \ket{1,0}_A \ket{0,0}_B$, in which a hard-core bosonic excitation on sublattice $B$ hops to sublattice $A$.
This is induced by the dipolar interaction, and occurs with a hopping matrix element $t^{AB}_{\b{r}} \sim e^{i \phi}$ with phase equal to the azimuthal angle $\phi$ between the dipoles.
This phase profile arises from the spherical harmonic $C^2_{-\Delta m = 1}(\theta,\phi)$, since the hopping changes the total angular momentum of the system by $\Delta m = -1$.
Sublattice B molecules are depicted as spinning to indicate their non-zero $z$-angular momentum in the excited state.
} 
\label{fig: hopping}
\end{figure}
}

\newcommand{\FigureTransitionOne}{
\begin{figure}[t]
\centering
\includegraphics[width=\columnwidth]{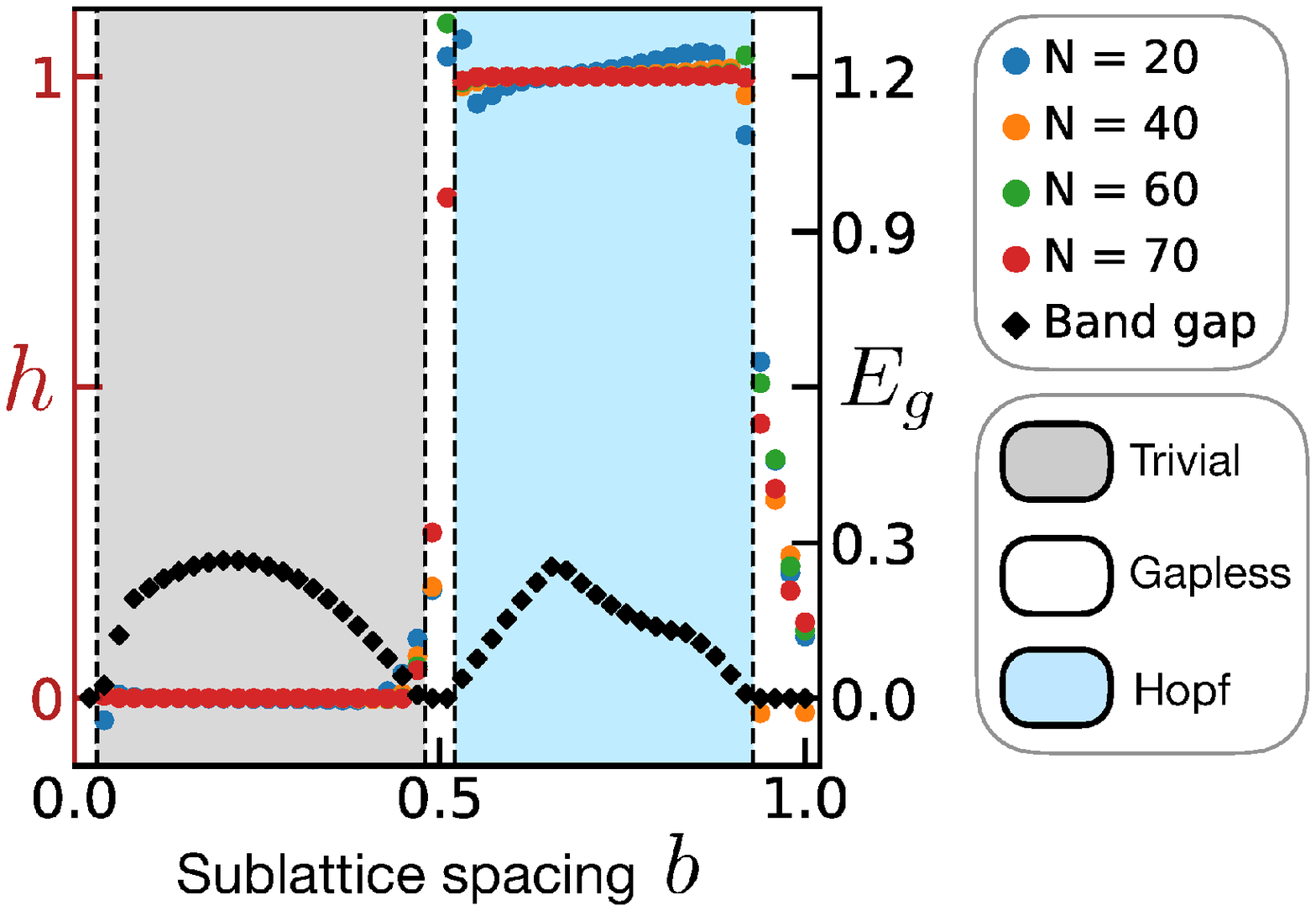}
\caption{
Numerical evaluation of the Hopf invariant $h$ for $N \times N \times N$ discretizations of momentum space (colored circles) and the band gap $E_g$ (black diamonds; in units of the nearest-neighbor hopping $t_{\text{nn}}$) of the specified dipolar spin system as a function of the vertical spacing $b$ between sublattices (in units of the nearest-neighbor spacing in the $xy$-plane), calculated with hopping range $R=8$.
The Hopf insulating phase (blue, right shaded) is observed across a large range of $b$; outside this range the system transitions to gapless (white) and trivial insulating (gray, left shaded) phases. Note that the Floquet modulation breaks the geometric symmetry $b \rightarrow 1 - b$, and hence the spectrum is not symmetric about $b = 0.5$.
} 
\label{fig: transition 2}
\end{figure}
}

\newcommand{\FigureTransitionTwo}{
\begin{figure}[t]
\centering
\includegraphics[width=\columnwidth]{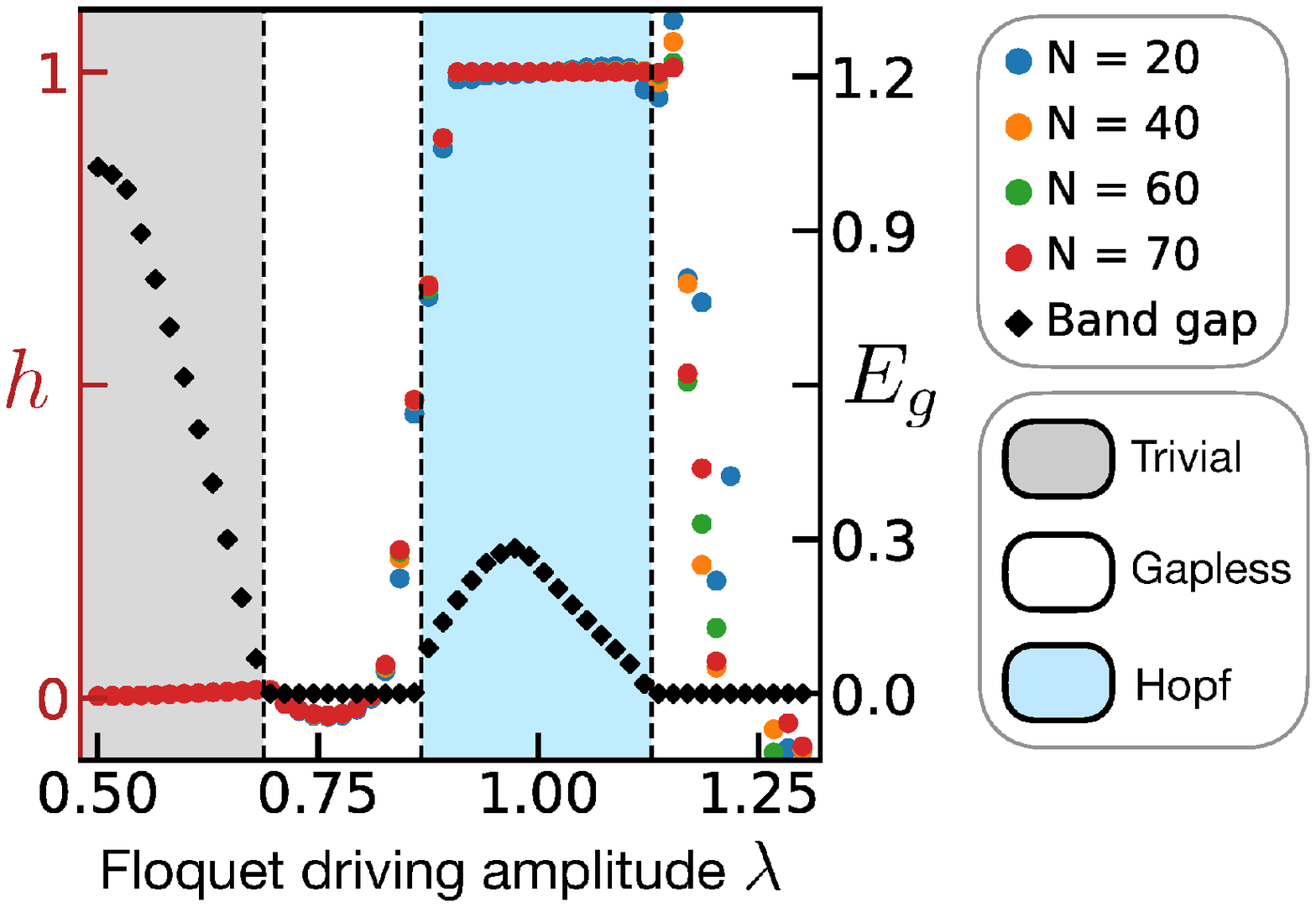}
\caption{
Numerical evaluation of the Hopf invariant $h$ for $N \times N \times N$ discretizations of momentum space (colored circles) and the band gap $E_g$ (black diamonds; in units of the nearest-neighbor hopping $t_{\text{nn}}$) of the specified dipolar spin system as a function of the strength $\lambda$ of the Floquet driving (defined in the main text), calculated with hopping range $R=8$.
The Hopf insulating phase (blue, right shaded) is observed across a range of $\lambda$; outside this range the system transitions to gapless (white) and trivial insulating (gray, left shaded) phases.
} 
\label{fig: transition 1}
\end{figure}
}

\newcommand{\FigureEdges}{
\begin{figure*}[t]
\centering
\includegraphics[width=\textwidth]{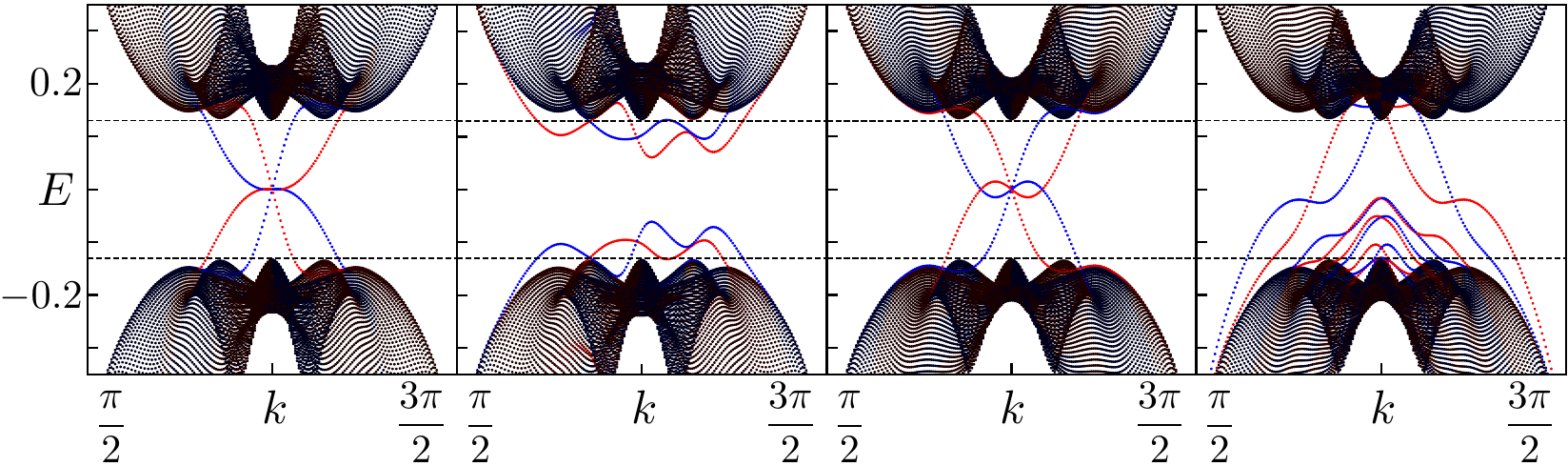}
\caption{
Spectra for the $(100)$-edge of the Hopf insulator along the diagonal $k_y = k, k_z = \pi + k$, calculated from the effective Hamiltonian in Eq.~(\ref{tony ham 0}).
Color indicates a mode's mean $x$-position, from red/light gray (localized at left edge), to black (bulk), to blue/dark gray (localized at right edge). Dashed lines mark the bulk band gap.
A sharp edge (open boundary conditions) respects the symmetry Eq.~(\ref{Cenke}) and leads to a gapless Dirac cone spectrum (far left). 
Adding a symmetry-breaking perturbation -- in this case, a chemical potential on the two sites nearest the edge -- gaps the Dirac cone (left center), demonstrating the non-adiabatic edge modes' lack of protection. 
In the adiabatic limit with edge termination smoothed over $\sim 20$ lattice sites, the edge spectrum is again gapless (right center).  
However, the Hopf invariant now protects the edge modes against \emph{all} smooth perturbations to edge, including a smoothed bump in the chemical potential in the edge region (far right).
For given transverse momenta $k_y, k_z$, the spectrum is calculated by first Fourier transforming the Hamiltonian along the $y$- and $z$-directions, and then performing exact diagonalization on the remaining 1D Hamiltonian.
Sharp/smooth edge spectra are calculated for a lattice with $80$/$160$ unit cells in the $x$-direction, and hoppings are truncated at a range $R=8$.
As a check on the high-frequency approximation leading to the effective Hamiltonian, Eq.~(\ref{tony ham 0}), we also perform the same computation for the \emph{exact} Floquet Hamiltonian, $H_F = i \log( \mathcal{T} \exp(-i \int_0^{2\pi/\Omega_{xy}} H(t) \, dt))$, at driving frequencies $\Omega_{xy} = 25 \, t_{\text{nn}}$, $\Omega_{z} = 600 \, t_{\text{nn}}$, and observe qualitatively identical edge spectra.
} 
\label{fig: edges}
\end{figure*}
}

\newcommand{\FigureExperimentFloquet}{
\begin{figure}[t]
\centering
\includegraphics[width=\columnwidth]{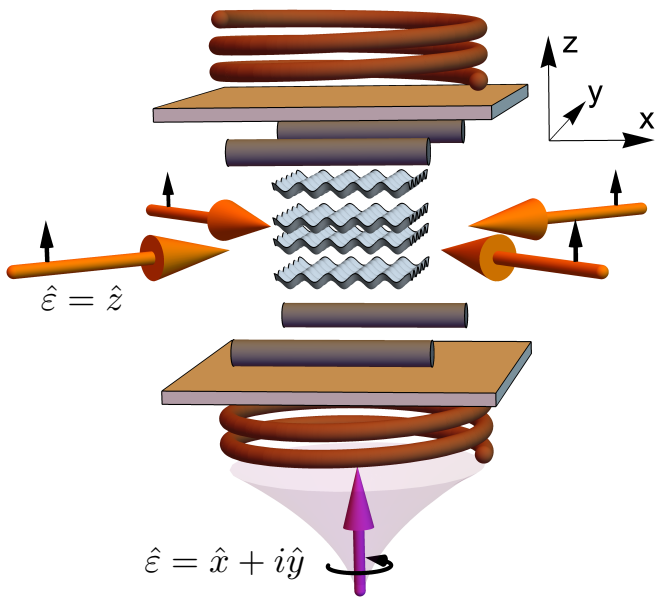}
\caption{
Schematic of the proposed experiment, highlighting the mechanisms for Floquet modulation.
The lattice (light gray waves) is formed by three standing waves of laser light (beams not pictured).
Stable electric field gradients are controlled an electrode system of tungsten rods (dark gray cylinders) and transparent plate electrodes (tan rectangles), while coils (brown spirals) generate a homogeneous magnetic field~\cite{covey2018enhanced}.
The $xy$-Floquet modulation is generated by $z$-polarized lasers forming a standing wave in the ($\hat{x} \pm \hat{y}$)-directions (large orange arrows, left and right; polarization in small black arrows), using the AC polarizability of ${}^{40}$K$^{87}$Rb.
The $z$-Floquet modulation is generated by a circularly-polarized laser in the $z$-direction (large purple arrow, bottom; polarization in small black arrow), which forms an intensity gradient along its direction of propagation due to the natural transverse spreading of a Gaussian laser beam.
}
\label{fig: experiment}
\end{figure}
}

\newcommand{\FigurePolarizability}{
\begin{figure*}[t]
\centering
\includegraphics[width=\textwidth]{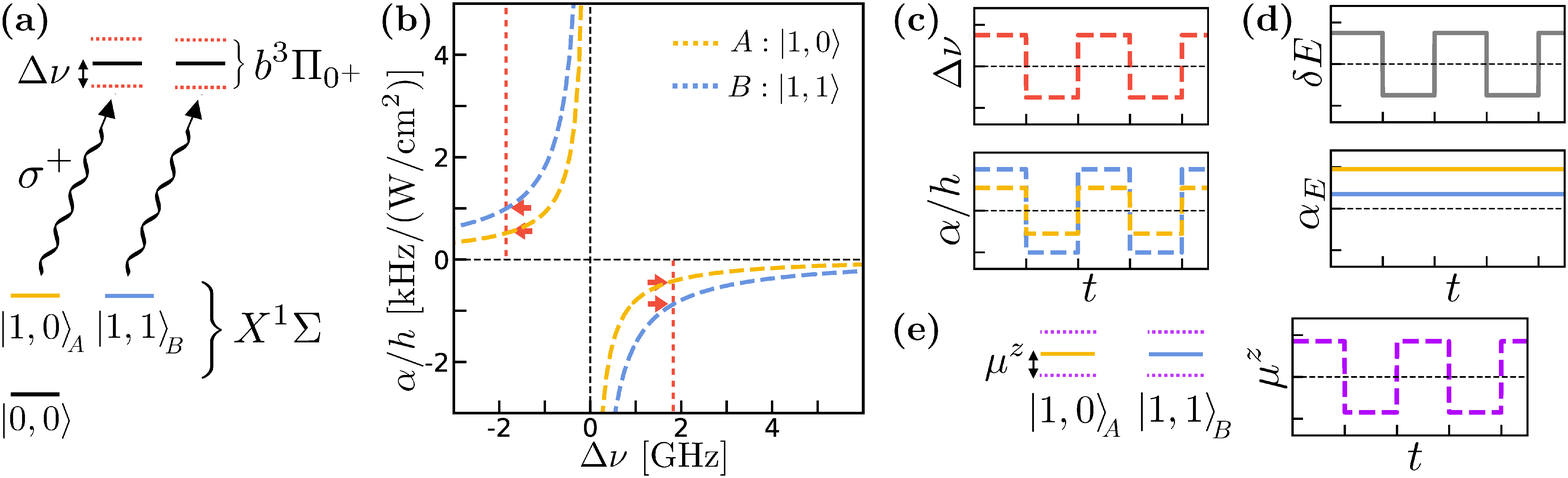}
\caption{
Depiction of the two-component driving scheme used to implement the $z$-gradient Floquet modulation.
$\textbf{(a)}$ One component is realized using circularly-polarized ($\sigma^+$) light tuned near, but off-resonant with the electronic transitions $|X^1\Sigma,v=0,J=1,m_J\rangle \to |b^3\Pi_{0^+},v=0,J=2,m_J+1\rangle$, with detuning $\Delta \nu$.
Here $X^1\Sigma,v=0$ denotes the electronic ground state manifold of the molecule, and $b^3\Pi_{0^+},v=0$ the relevant electronic excited state manifold.
This induces energy shifts in the electronic ground states of interest, $\ket{1,0}_A$ and $\ket{1,1}_B$, proportional to the AC polarizability $\alpha$ of ${}^{40}$K$^{87}$Rb at the particular detuning.
$\textbf{(b)}$ AC polarizabilities under circularly polarized $\sigma^+$ light as a function of the detuning $\Delta\nu$, calculated from first principles (see the Supplemental Material for details). 
Red dotted lines label two detunings that are oscillated between to achieve a step function Floquet modulation. Red arrows indicate the corresponding polarizibilities.
$\textbf{(c)}$ Simplified depiction of the detuning and resulting polarizibilities as a function of time $t$.
In the dipolar simulations, we use a higher parameter step function, Eq.~(\ref{step function}), which allows greater flexibility to optimize the band gap of the Hopf insulating phase.
The AC light intensity is held constant in time (not depicted).
$\textbf{(d)}$ The other component of modulation consists of an electric field gradient $\delta E$ oscillated in time according to the same step function.
The (DC) polarizibilities $\alpha_E$ of the $\ket{1,0}_A$, $\ket{1,1}_B$ states under this field are constant in time.
$\textbf{(e)}$ The polarizibilities and field amplitudes in $\textbf{(c-d)}$ multiply to produce oscillating energy shifts $\mu^z$ (dotted purple) of the $\ket{1,0}_A$, $\ket{1,1}_B$ states.
While each individual component of the $z$-gradient modulation produces a different magnitude shift for each state owing to the states' differing polarizibilities, the linear combination of both components can be chosen to produce equal shifts.
}
\label{fig: polarizability}
\end{figure*}
}

\begin{abstract}

We present a quantitative, near-term experimental blueprint for the quantum simulation of topological insulators using lattice-trapped ultracold polar molecules. 
In particular, we focus on the so-called Hopf insulator, which represents a three-dimensional topological state of matter existing outside the conventional tenfold way and crystalline-symmetry-based classifications of topological insulators.
Its topology is protected by a \emph{linking number} invariant, which necessitates long-range spin-orbit coupled hoppings for its realization.
While these ingredients have so far precluded its realization in solid state systems and other quantum simulation architectures, in a companion manuscript~\cite{schuster2019realizing} we predict that Hopf insulators can in fact arise naturally in dipolar interacting systems. 
Here, we investigate a specific such architecture in lattices of polar molecules, where the effective `spin' is formed from sublattice degrees of freedom.
We introduce two techniques that allow one to optimize dipolar Hopf insulators with large band gaps, and which should also be readily applicable to the simulation of other exotic bandstructures.
First, we describe the use of Floquet engineering to control the range and functional form of dipolar hoppings and second, we demonstrate that molecular AC polarizabilities (under circularly polarized light) can be used to precisely tune the resonance condition between different rotational states. 
To verify that this latter technique is amenable to current generation experiments, we calculate from first principles the AC polarizability for $\sigma^+$ light for ${}^{40}$K$^{87}$Rb.
Finally, we show that experiments are capable of detecting the unconventional topology of the Hopf insulator by varying the termination of the lattice at its edges, which gives rise to three distinct classes of edge mode spectra.

\end{abstract}

\maketitle

%
%



The rich internal structure of ultracold polar molecules has led to intense interest for their use in a wide range of applications, ranging from quantum simulation and  computation, to ultracold chemistry and precision measurement~\cite{sage2005optical,park2015ultracold,takekoshi2014ultracold,guo2016creation,ciamei2018rbsr,molony2014creation,tung2013ultracold,deiglmayr2010permanent,kozlov1995parity,hudson2006cold,baron2014order,safronova2018search,balakrishnan2016perspective,ni2018dipolar,sawant2020ultracold,ni2008high}.
Understanding and controlling this structure has led to the development of a host of techniques enabling the preparation and manipulation of rovibrational states in polar molecules~\cite{ni2008high,moses2017new,aldegunde_hyperfine_2008,aldegunde_manipulating_2009,ospelkaus_controlling_2010,yan_observation_2013,moses2015creation,aldegunde_hyperfine_2008,ospelkaus_controlling_2010,neyenhuis_anisotropic_2012,de2019degenerate,anderegg2019optical}. 
From the perspective of quantum simulation, polar molecules enjoy a unique advantage compared to their neutral atom cousins, owing to the presence of strong, anisotropic, long-range dipolar interactions; these interactions have proven useful for theoretical proposals aiming to realize a number of exotic phases, including disordered quantum magnets~\cite{hazzard2013far,gorshkov2013kitaev,yao_quantum_2018}, Weyl semimetals \cite{syzranov2016emergent} and fractional Chern insulators~\cite{yao2013realizing,yao_quantum_2018}. 
Motivated, in part, by these prospects, the last decade has seen tremendous experimental progress, advancing from rovibrational ground state cooling~\cite{ni2008high} to the recent realization of a Fermi degenerate molecular gas~\cite{de2019degenerate}.
Moreover, from a geometric perspective, molecules can either be loaded into optical lattices~\cite{moses2015creation} or optical tweezer arrays~\cite{anderegg2019optical}.
%
As in other quantum simulation platforms, Floquet engineering~\cite{bukov2015universal,lee_floquet_2016} -- high-frequency, periodic time-modulation -- can further sculpt the molecules' interaction, broadening the scope of accessible phases~\cite{micheli2007cold,lee2018floquet}.

In this article, we provide an explicit experimental blueprint for realizing another heretofore unobserved phase of matter, the Hopf insulator, in polar molecules.
The Hopf insulator is a particular topological insulator, characterized by a \emph{linking} number topological invariant arising from the unique topology of knots in three dimensions and the Hopf map of mathematics~\cite{hopf1931abbildungen,moore_topological_2008}.
Notably, it exists only in two-band systems, falling outside the traditional `tenfold way' classification of topological insulators~\cite{schnyder_classification_2008,kitaev_periodic_2009} and suggesting that it might possess different physics than the most well-known examples of these phases.
Despite much interest in both the Hopf insulator~\cite{moore_topological_2008,deng_hopf_2013,deng_systematic_2014,deng_probe_2016,kennedy_homotopy_2015,kennedy_topological_2016,liu_symmetry_2017,yuan2017observation,alexandradinata2019actually,schuster2019floquet,he2019three,he2020non,hu2020quench} and physics associated with the Hopf map more generally~\cite{ackerman2017static,wang_scheme_2017,tarnowski2017characterizing,yan_nodal-link_2017}, experimentally realizing the Hopf insulator has remained an open challenge, and even proposed  implementation platforms (e.g.~in either conventional quantum materials or cold atomic quantum simulators) remain few and far between \cite{deng_probe_2016,yuan2017observation}. 
The key challenges arise directly from the nature of the Hopf map. In particular, realizing the Hopf insulator requires two essential ingredients: 1) the presence of long-range hoppings and 2) strong spin-orbit coupling, manifested in hoppings whose phase is spatially anisotropic.

In a companion manuscript~\cite{schuster2019realizing}, we predict that combining the dipolar interaction with Floquet engineering~\cite{bukov2015universal,lee_floquet_2016} can naturally give rise to the Hopf insulator in interacting spin systems.
Here, we build upon this result by providing a quantitative blueprint using lattice-trapped ultracold polar molecules, focusing for concreteness on   $^{40}$K$^{87}$Rb~\cite{ni2008high,moses2015creation,yan_observation_2013,ospelkaus_controlling_2010,aldegunde_manipulating_2009,aldegunde_hyperfine_2008}.
Our approach takes advantage of the full toolset of controls developed for polar molecular systems.
In particular, we envision a deep, three-dimensional optical lattice, so that the molecules' rotational motion constitutes the fundamental degrees of freedom in the system.
Rotational excitations are exchanged between lattice sites via the dipolar interaction, which simulates the hopping of hardcore bosons  on the lattice. 
The two band, or `spin', degrees of freedom of the Hopf insulator are formed from two sublattices, distinguished from each other by the lattice light itself -- different intensity light forming the two sublattices induces different level structures in the trapped molecules, according to the molecules' polarizability~\cite{neyenhuis_anisotropic_2012}.

In contrast to prior studies \cite{yao_quantum_2018,peter_topological_2015,yao2013realizing}, we utilize this polarizability to isolate the $\Delta m = \pm 1$ angular-momentum-changing component of the dipolar interaction, which \textit{precisely} induces the requisite spin-orbit coupling of the Hopf insulator~\cite{moore_topological_2008}. 
To complete our construction, we demonstrate that Floquet engineering can be implemented using amplitudes of applied laser light and DC electric fields which are easily accessible in current generation experiments; moreover, we show that this engineering can tune the system's hoppings into the Hopf insulating phase with large band gaps $\gtrsim 0.26 \, t_{\text{nn}}$ (in units of the nearest-neighbor hopping, $t_{\text{nn}}$), enabling easier experimental observation.
Finally, a particularly simple way to achieve the requisite rotational level structure (Fig.~1) is to utilize circularly-polarized optical radiation in conjunction of the molecule's AC polarizability. To this end, in order to demonstrate quantitative feasibility, we provide the first detailed calculations of the relevant circular polarizabilities for $^{40}$K$^{87}$Rb. 

Direct experimental verification of the Hopf insulator is most simply achieved through spectroscopy of its gapless edge modes.
In a companion manuscript~\cite{schuster2019realizing}, we demonstrate that these edge modes are robust at any smooth boundary of the Hopf insulating phase, while for sharp boundaries their presence or absence signifies the existence of an underlying crystalline symmetry~\cite{liu_symmetry_2017}.
We will show that all three of these qualitatively distinct boundary spectra can be manufactured and probed in ultracold polar molecule simulations.
Since the Hopf insulator's edge behavior is a direct result of it being outside the conventional tenfold way, this serves as a direct experimental probe of the Hopf insulator's unique topological classification.

Our manuscript is structured as follows.  We begin with an overview of the Hopf insulator, with a specific focus on the requirements -- a two band system, and long-range, spin-orbit coupled hoppings.
We then turn to the setting of our proposal, outlining precisely how the rotational excitations of polar molecules can simulate spin-orbit coupled particles hopping on a lattice.
Next, we demonstrate how particular patterns of Floquet driving can provide tremendous control over these hoppings, and numerically verify that these can be used to tune the system into a large band-gap, Hopf insulator phase.
We present the edge modes of the polar molecular Hopf Hamiltonian, and show that they display three qualitatively distinct spectra dependent on the lattice termination.
Finally, we conclude by providing a detailed description of all aspects of the proposal's implementation in a three dimensional optical lattice  of $^{40}$K$^{87}$Rb.

\FigureLattice

\section{The Hopf Insulator}

We begin with an introduction to the Hopf insulator, seeking to motivate the connection between the linking number interpretation of the Hopf invariant and the long-range spin-orbit coupling required for its physical realization. 

The Hopf insulator is a particular type of topological insulator~\cite{thouless1982quantized,haldane1988model,kane2005z,konig2007quantum,fu2007topological,moore2007topological,roy2009topological,zhang2009topological}, a class of phases of matter most notable for exhibiting conducting surface states despite an insulating bulk. 
They are differentiated from conventional insulators by a non-zero topological invariant associated with their underlying spin-orbit-coupled band structure; moreover, their surface states are unusually robust to the detrimental effects of impurities.
Their organization was first captured via the so-called ten-fold way classification~\cite{schnyder_classification_2008,kitaev_periodic_2009}, and consists of a wide landscape of phases dependent on a system's dimensionality and symmetries.
Nevertheless, more recent work has exposed topological insulators that exist beyond this classification framework; notable examples include topological crystalline insulators~\cite{fu_topological_2011}, higher-order topological insulators~\cite{schindler2018higher}, and our insulator of interest, the Hopf insulator~\cite{moore_topological_2008,deng_hopf_2013,deng_systematic_2014,deng_probe_2016,kennedy_topological_2016,yuan2017observation,liu_symmetry_2017}.

The Hopf insulator exists in three-dimensions in the \emph{absence} of any symmetries, for which the ten-fold way classification~\cite{schnyder_classification_2008,kitaev_periodic_2009} would nominally predict only an ordinary insulator.  
In our context, it will consist of hard-core boson degrees of freedom hopping on a three-dimensional lattice (although one is accustomed to thinking of topological insulators in terms of fermions, their single-particle nature also enables a hard-core bosonic realization).
The bosons come in two `pseudospins', $A$ and $B$, which will form the two bands of the system. 
These may be formed from physical spins, but are not required to be -- in our realization, they will correspond to two sublattices of the three-dimensional lattice.
In real space, the Hopf insulator Hamiltonian takes the generic form
\vspace{-.05cm}
\begin{equation}\label{Htb}
H_{\text{eff}} =  \! \frac{1}{2} \!\! \sum_{\substack{\b{v}, \b{r} \neq \b{0}, \\ \alpha, \beta}}\!\! \big[ t^{\alpha \beta}_{\b{r}} a^{\dagger}_{\b{v} + \b{r}, \alpha} a_{\b{v}, \beta} + \textit{h.c.} \big] + \! \sum_{\b{v}, \alpha} \mu^{\alpha}  a^{\dagger}_{\b{v}, \alpha} a_{\b{v}, \alpha},
\vspace{-.2cm}
\end{equation}
where $a^{\dagger}_{\b{v}, \alpha}$ is the creation operator for a hard-core boson at lattice site $\b{v}$ of pseudospin $\alpha  \in \{ A, B \}$.
The Hamiltonian consists of both pseudospin-preserving ($t^{AA}_{\b{r}}$ and $t^{BB}_{\b{r}}$) and pseudospin-flipping ($t^{AB}_{\b{r}}$ and $t^{BA}_{\b{r}}$) hoppings, as well as a pseudospin-dependent chemical potential $\mu^{\alpha}$. 

\FigureHopf

The topology of the Hopf insulator is most easily seen in its momentum-space representation, governed by the two-by-two matrix $H^{\alpha \beta}(\b{k}) = \sum_{\b{r}} \tilde{t}^{\alpha \beta}_{\b{r}} e^{i \b{k} \cdot \b{r}} + \mu^\alpha \delta^{\alpha \beta}$.
This is conveniently decomposed as
$H(\b{k}) = n_0(\b{k}) \mathbbm{1} +  \b{n}(\b{k}) \cdot \boldsymbol{\sigma}$,
where the Pauli matrices $\boldsymbol{\sigma}$ act on the pseudospin degrees of freedom, which form the two bands of the Hopf insulator, and the condition that the bands are gapped requires $|\b{n}(\b{k})| > 0$.
We can view this Hamiltonian as a map that takes vectors $\b{k}$ in the Brillouin zone to points $\hat{\b{n}} \equiv \b{n}/|\b{n}|$ on the Bloch sphere.
To see the Hopf insulator's topology, consider the \emph{pre-images} of two different Bloch sphere points $\hat{\b{n}}, \hat{\b{n}}'$ in the Brillouin zone, i.e. the set of momenta $\b{k}$ such that $\hat{\b{n}}(\b{k}) = \hat{\b{n}}$, or $\hat{\b{n}}(\b{k}) = \hat{\b{n}}'$. Since the Brillouin zone is three-dimensional -- one dimension higher than the Bloch sphere -- these pre-images are generically 1D loops in the Brillouin zone. The Hopf invariant $h$ of the Hamiltonian $H(\b{k})$ is precisely equal to the \emph{linking number} of these two loops, for any choice of $\hat{\b{n}}, \hat{\b{n}}'$ [Fig.~\ref{fig: hopf}(a)]. The invariant can be calculated from the Bloch Hamiltonian via the Chern-Simons form~\cite{moore_topological_2008}: 
\begin{equation}\label{invariant formula}
h = \int_{BZ} d^3\b{k} \,\,  j^{\mu}(\b{k}) A_{\mu}(\b{k}),
\end{equation}
where $ j^\mu(\b{k}) = \frac{1}{8\pi} \epsilon^{\mu \nu \lambda} \hat{\b{n}} \cdot (  \partial_{k_\nu} \hat{\b{n}} \times  \partial_{k_\lambda} \hat{\b{n}} )$
is the  Berry curvature and $A_\mu(\b{k})$ its associated vector potential.

The linking number interpretation leads to two observations, one which explains the need for long-range hoppings and the other which justifies the required form of spin-orbit coupling. 
First, the rapid variation in $\b{n}(\b{k})$ required for pre-image linking necessitates the presence of strong long-range hoppings, which contribute oscillations $\sim \!e^{i \b{k} \cdot \b{r}}$ to $\b{n}(\b{k})$, at a frequency proportional to their range $\b{r}$. 
Specifically, no nearest neighbor Hamiltonian is known for the Hopf insulator; the prototypical Hopf insulator Hamiltonian~\cite{moore_topological_2008} features as far as next-next-nearest neighbor hoppings.
Second, pre-image linking by definition requires a strong coupling between the pseudospin degree of freedom and the momentum, much as is true for other topological insulators.
Inspired by the model of Ref.~\cite{moore_topological_2008}, in this work we realize a specific form of this spin-orbit coupling, generated via pseudospin-flipping hoppings with a direction-dependent phase $t^{AB}_{\b{r}} \sim e^{i \phi}$, where $\phi$ is the azimuthal angle of the hopping displacement $\b{r}$ (Fig.~\ref{fig: hopping}).
This form of hopping locks the $n_x, n_y$ components of pseudospin to the $k_x, k_y$ components of the momentum, such that the pre-image of, e.g. $\hat{\b{n}} = \hat{\b{x}}$, is exactly a 90 degree rotation about the $k_z$-axis of the pre-image of $\hat{\b{n}} = \hat{\b{y}}$.
As illustrated in Fig.~\ref{fig: hopf}, this simple correspondence leads naturally to linking of the two pre-images.
While this simple argument applies only to pre-images related by 90 or 180 degree rotations about the $z$-axis (due to the cubic lattice symmetry), this is in fact sufficient: in a gapped model, the linking number is \emph{constant} for all pairs of pre-images.
We note that this same phase profile of the hoppings is also present in two-dimensional realizations of Chern insulating physics, both in the prototypical Qi-Wu-Zhang model~\cite{qi2006topological} as well as in positionally disordered systems~\cite{agarwala2019topological}

In the following two sections, we demonstrate that systems of dipolar interacting spins provide a natural ground to realize both of these key ingredients.
We begin by describing how a particular configuration of the spins' level structures leads to the effective hard-core boson Hamiltonian of Eq.~(\ref{Htb}), including the desired spin-orbit coupling $t^{AB}_{\b{r}} \sim e^{i \phi}$.
We then augment the bare dipolar hoppings with a Floquet engineering scheme, which serves to decrease the relative strength of nearest-neighbor hoppings and provides useful experimental parameters for tuning into the Hopf insulating phase.

\section{The Dipolar Hamiltonian}\label{sec: dipolar}

We now turn to the setting of our proposal. 
We envision a three-dimensional optical lattice filled with ultracold polar molecules. 
We work in the deep lattice limit, so that the molecules themselves do not hop between lattice sites, and the molecules' rotational states form the fundamental degrees of freedom of our system~\cite{yan2013realizing}. 
As shown in Fig.~\ref{fig: lattice}, the lattice is formed by alternating planes of two-dimensional square lattices, stacked in the $z$-direction. These form two sublattices, $A$ and $B$, which will play the role of the pseudospin in the Hopf insulator.

The molecules are most strongly governed by the rotational Hamiltonian $H_{\text{rot}} = \Delta \b{J}^2$, with eigenstates $\ket{J,m_J}$ indexed by their orbital ($J$) and magnetic ($m_J$) angular momentum quantum numbers, which have energies $E = \Delta J(J+1)$ and wavefunctions described by the spherical harmonic functions~\cite{fn2}. 
While naturally organized into degenerate manifolds of each $J$, the $m_J$ eigenstates are split by both intrinsic hyperfine interactions and \emph{tunable} extrinsic effects resulting from electric fields, magnetic fields and incident laser light. 
These extrinsic effects (which set the molecules' quantization axis, i.e.~$\hat{\b{z}}$ in Fig.~\ref{fig: lattice}) enable a direct modulation of the rotational states' energies in both \emph{space}  (to distinguish between the $A$ and $B$ sublattices) and  \emph{time} (to implement Floquet engineering).

We now aim to use these rotational states to realize an effective Hamiltonian of hard-core bosons, as in Eq.~(\ref{Htb}).
We focus on the lowest four rotational eigenstates (i.e. the $J=0,1$ manifolds), and use these to define two distinct hard-core bosonic degrees of freedom.
On the $A$-sublattice we form a hard-core boson from the doublet $\{ |0_A \rangle = \ket{0,0}_A, |1_A \rangle = \ket{1,0}_A \}$, while on the $B$-sublattice we utilize $\{ |0_B \rangle = \ket{0,0}_B, |1_B \rangle = \ket{1,1}_B \}$, as illustrated in Fig.~\ref{fig: lattice}. 
The hard-core bosons interact with each other through the dipolar interaction~\cite{brown2003rotational}:
\begin{equation}\label{Hdd}
H^{ij}_{\text{dd}} = \frac{-\sqrt{6}}{4 \pi \epsilon_0 r^3}  \!\!\! \sum_{\Delta m_J=-2}^{2}  \!\!\!\!\! C^2_{- \Delta m_J}(\theta,\phi) T^2_{\Delta m_J}(\b{d}^{(i)},\b{d}^{(j)}),
\end{equation}
where $(r, \theta, \phi)$ parameterizes the separation of the interacting molecules $i$ and $j$ in spherical coordinates, and we compress unit and sublattice indices into a single index $i = \b{v},\alpha$. The dipole moment operator $\b{d}^{(i)} = (d_-^{(i)}, d_z^{(i)} d_+^{(i)}) $ is a rank-1 spherical tensor acting on the rotational states of the molecule $i$, whose three components change the molecule's magnetic quantum number by $(-1,0,+1)$ respectively. The spherical harmonics $C^2_{-\Delta m}(\theta,\phi)$ capture the spatial dependence of the interaction, and are accompanied by the corresponding component of $T^2_{\Delta m}$, the unique rank-2 spherical tensor generated from the dipole operators $\b{d}^{(i)}$, $\b{d}^{(j)}$. Explicitly, we have $T^2_{\pm 2} = d_{\pm}^{(i)} d_{\pm}^{(j)}$, $T^2_{\pm 1} = (d_{\pm}^{(i)} d_{z}^{(j)} + d_{z}^{(i)} d_{\pm}^{(j)})/\sqrt{2}$, $T^2_{0} = (d_{\pm}^{(i)} d_{\mp}^{(j)} + 2d_{z}^{(i)} d_{z}^{(j)} + d_{\mp}^{(i)} d_{\pm}^{(j)})/\sqrt{6}$. 

\FigureHopping

A few remarks are in order. 
First, we will assume that the dipolar interaction strength is significantly weaker than the energy splittings within the $J=1$ manifold. 
Second, we will tune the splitting between the $|0_A \rangle$ and $ |1_A \rangle$ states to be resonant with that of the $|0_B\rangle$ and $ |1_B\rangle$ states (Fig.~\ref{fig: lattice}).
Conservation of energy then dictates that the dipolar interaction can only induce transitions within our prescribed hard-core bosonic doublets, i.e.~those that preserve boson number. These transitions take the form of hoppings in the bosonic Hamiltonian, $t_{ij} = \bra{0_i, 1_j} H^{ij}_{\text{dd}} \ket{1_i, 0_j}$.
These hoppings may occur either within a sublattice ($t^{AA}_{\b{r}}$ and $t^{BB}_{\b{r}}$) or across sublattices ($t^{AB}_{\b{r}}$).
With the prescribed geometry and level structure, we have:
\begin{equation}\label{tSL}
\begin{split}
t^{AA}_{\b{r}} & = - \frac{d_{00}^2}{4 \pi \epsilon_0} \,  \frac{3\cos^2(\theta) - 1}{r^3} \\
t^{BB}_{\b{r}}  &  = \frac{d_{01}^2}{8 \pi \epsilon_0} \, \frac{3\cos^2(\theta) - 1}{r^3} \\
t^{AB}_{\b{r}}  = (t^{BA}_{-\b{r}})^* &  = -\frac{3 \, d_{00} d_{01}}{4 \sqrt{2} \pi \epsilon_0} \, \frac{\cos(\theta)\sin(\theta)}{r^3} \, e^{i\phi}, \\
\end{split}
\end{equation}
where $(r,\theta,\phi)$ parameterizes the displacement between sites in spherical coordinates, equal to $\b{r}$ for intra-sublattice hoppings and $\b{r} + b \, \hat{\b{z}}$ for inter-sublattice hoppings (where $b$ is the vertical distance between $A$ and $B$ planes), and $d_{00}$, $d_{01}$ are the dipole moments $d_{00} = \bra{1,0} d_z \ket{0,0}$ and $d_{01} = \bra{1,\pm 1} d_{\pm} \ket{0,0}$. 
Our choice of rotational states guarantees that the inter-sublattice hopping, $t^{AB}_{\b{r}}$, arises solely from the $\Delta m_J = +1$ term in $H_{\textrm{dd}}$, which gives rise to a hopping phase $t^{AB}_{\b{r}} \sim e^{i \phi}$ via the $C^2_{- 1}$ spherical harmonic. 
As illustrated in Fig.~\ref{fig: hopf}, this naturally leads to linking between the Bloch sphere pre-images.
Finally, variations in the energy splitting between sublattices naturally appear as effective chemical potentials $\mu^\alpha$, completing the realization of the Hamiltonian Eq.~(\ref{Htb}).

\section{Floquet engineering}\label{sec: floquet}

While the dipolar interaction elegantly realizes the requisite spin-orbit coupling, relatively strong nearest neighbor hopping as well as the slow asymptotic decay of the  $1/R^3$ power-law preclude numerical observation of  Hopf insulating behavior. 
To this end, we utilize Floquet engineering to two effects: first, to decrease `odd' hoppings in the $xy$-plane (those with odd $r_x + r_y$) and second, to truncate the dipolar power-law in the $z$-direction~\cite{lee_floquet_2016}.
We achieve each effect by adding spatio-temporal dependence to the chemical potential $\mu^\alpha_\b{v}(t)$, and oscillating each $\mu^\alpha_\b{v}(t)$ at timescales much faster than the hopping.
Under certain conditions (specified below), this leads to an effective time-\emph{independent} Hamiltonian of the same form as Eq.~(\ref{Htb}), but with modified hoppings
\begin{equation}
t^{\alpha \beta}_{\b{r}} \rightarrow \beta^{\alpha \beta}_{\b{r}} \, t^{\alpha \beta}_{\b{r}},
\end{equation}
where the damping coefficients, $\beta^{\alpha \beta}_{\b{r}}$, are determined by the specific profiles of the oscillated chemical potentials, $\mu^\alpha_\b{v}(t)$.
In what follows, we first derive this relation explicitly [Eq.~(\ref{beta gen 2})], and then introduce two Floquet engineering schemes [i.e. explicit profiles for the spatio-temporal dependence of $\mu^\alpha_\b{v}(t)$] that achieve the hopping modifications described above.

\subsubsection{Overview of Floquet engineering}

We begin with a broad introduction to Floquet engineering using a time-dependent chemical potential, following Ref.~\cite{lee_floquet_2016} but modified to include sublattices and complex hoppings.
We consider a time-dependent Hamiltonian of the form Eq.~(\ref{Htb}) where the chemical potential $\mu^\alpha_\b{v}(t)$ now varies with the lattice site $\b{v}$ as well as periodically in time $t$, with a period $T$.
To calculate the effect of the driving, we move into a rotating frame, defining the unitary
\begin{equation}\label{rot U}
U(t) = \exp \bigg[ - i \int_0^t dt' \, \frac{1}{2} \sum_{\b{v},\alpha} \mu^{\alpha}_{\b{v}}(t) \sigma^z_{\b{r}}   \bigg],
\end{equation}
and the rotated wavefunction
\begin{equation}\label{rot wavefn}
\ket{ \Psi'(t)} = U^{\dagger}(t) \ket{ \Psi(t)},
\end{equation}
whose time-evolution is governed by the Hamiltonian
\begin{equation}\label{rot H}
\begin{split}
H'(t) = & U^{\dagger}(t) H(t) U(t) - i U^{\dagger}(t) \dot{U}(t) \\
 = & \frac{1}{2} \sum_{\substack{\b{v}, \b{r} \neq \b{0}, \\ \alpha, \beta}} \bigg( \exp \bigg[  -i \int_{0}^{t} dt' \,  (\mu^{\alpha}_{\b{v}+\b{r}}(t') - \mu^{\beta}_{\b{v}}(t')) \bigg] \times \\ 
& t^{\alpha \beta}_{\b{r}} \, a^{\dagger}_{\b{v} + \b{r}, \alpha} a_{\b{v}, \beta} + \textit{h.c.} \bigg).  \\
\end{split}
\end{equation}

At high-frequencies, $1/T \gg | t^{\alpha \beta}_{\b{r}} |$, the rotated Hamiltonian is well-approximated by replacing all quantities by their average over a single period. This gives an effective time-independent Hamiltonian
\begin{equation}\label{tony ham 0}
\begin{split}
H^{\text{eff}} = & \frac{1}{2} \sum_{\substack{\b{v}, \b{r} \neq \b{0}, \\ \alpha, \beta}} \big[ \beta^{\alpha \beta}_{\b{v}+\b{r},\b{v}} \, t^{\alpha \beta}_{\b{r}} \, a^{\dagger}_{\b{v} + \b{r}, \alpha} a_{\b{v}, \beta} + \textit{h.c.} \big] \\
& + \sum_{\b{v}, \alpha} \mu^{\alpha}_{\b{v}} \, a^{\dagger}_{\b{v}, \alpha} a_{\b{v}, \alpha},
\end{split}
\end{equation}
with a static chemical potential 
\begin{equation}
\mu^{\alpha}_{\b{v}} = \frac{1}{T} \int_{0}^{T} dt \, \mu^{\alpha}_{\b{v}}(t),
\end{equation}
and hoppings suppressed by the dampings
\begin{equation}\label{beta gen 1}
\begin{split}
\beta^{\alpha \beta}_{\b{v}+\b{r},\b{v}} =&  \frac{1}{T} \int_{0}^{T} dt \, \exp \bigg[  -i \int_{0}^{t} dt' \,  (\mu^{\alpha}_{\b{v}+\b{r}}(t') - \mu^{\beta}_{\b{v}}(t')) \bigg] \\
=&  \frac{1}{T} \int_{0}^{T} dt \, \bigg( \cos \bigg[  \int_{0}^{t} dt' \, (\mu^{\alpha}_{\b{v}+\b{r}}(t') - \mu^{\beta}_{\b{v}}(t')) \bigg] \\
& - i \sin \bigg[  \int_{0}^{t} dt' \, (\mu^{\alpha}_{\b{v}+\b{r}}(t') - \mu^{\beta}_{\b{v}}(t')) \bigg] \bigg). \\
\end{split}
\end{equation}
For convenience, we will always choose $\mu^{\alpha}_{\b{v}}(t)$ to be an even function of $t$, in which case the imaginary part of the damping vanishes and we have
\begin{equation}\label{beta gen 2}
\begin{split}
\beta^{\alpha \beta}_{\b{v}+\b{r},\b{v}} & = \frac{1}{T} \int_{0}^{T} dt \, \cos \bigg[  \int_{0}^{t} dt' \, (\mu^{\alpha}_{\b{v}+\b{r}}(t') - \mu^{\beta}_{\b{v}}(t')) \bigg].  \\
\end{split}
\end{equation}
In this case, the dampings modulate only the hoppings' magnitudes, and not their phase.

Since the modulation is generically inhomogeneous, care must be taken to ensure that the dampings are in fact translation invariant, $\beta^{\alpha \beta}_{\b{v}+\b{r},\b{v}} = \beta^{\alpha \beta}_{\b{r}}$, if one desires translation invariance in the effective Hamiltonian.
This constraint requires that $\cos \big[  \int_{0}^{t} dt' \, (\mu^{\alpha}_{\b{v}+\b{r}}(t') - \mu^{\beta}_{\b{v}}(t')) \big]$ be independent of $\b{v}$. For intra-sublattice hoppings ($\alpha = \beta$), there are two ways to achieve this: 1) with a `gradient' modulation, where $\mu^{\alpha}_{\b{v}}(t)$ is linear in $\b{v}$, and 2) with an `even-odd' modulation $\mu^{\alpha}_{\b{v}}(t) = \mu^{\alpha} (-1)^{s_x v_x + s_y v_y + s_z v_z}, \, s_i \in \{ 0 , 1\}$. (The latter is possible because we restrict to the cosine term of Eq.~(\ref{beta gen 1}), which is even in $\mu$ and thus requires only the absolute value of $\mu^{\alpha}_{\b{v}+\b{r}}(t') - \mu^{\alpha}_{\b{v}}(t')$ to be independent of $\b{v}$.) For inter-sublattice hoppings ($\alpha \neq \beta$), this constraint additionally requires that the sublattices' modulations differ only by a \emph{position-independent} function of time, namely
\begin{equation}\label{Floquet mu form}
\begin{split}
\mu^A_{\b{v}}(t) = \mu_{\b{v}}(t) \\
\mu^B_{\b{v}}(t) = \mu_{\b{v}}(t) + \mu_{SL}(t).
\end{split}
\end{equation}
These lead to damping coefficients
\begin{equation}\label{beta AB}
\begin{split}
\beta^{AA}_{\b{r}} & = \beta^{BB}_{\b{r}} = \frac{1}{T} \int_{0}^{T} dt \, \cos \bigg[  \int_{0}^{t} dt' \, (\mu_{\b{v}+\b{r}}(t') - \mu_{\b{v}}(t')) \bigg] \\
\beta^{AB}_{\b{r}} & = \!\! \frac{1}{T} \!\! \int_{0}^{T} \!\!\! dt \,  \cos \bigg[  \int_{0}^{t} dt' \, (\mu_{\b{v}+\b{r}}(t') - \mu_{\b{v}}(t') - \mu^{SL}(t')) \bigg]  \\
\end{split}
\end{equation}
for the intra- and inter-sublattice hoppings. 
We must also ensure that $\mu^{\alpha}_{\b{v}}$ is translation invariant, which requires only that the average modulation is the same in each unit cell $\b{v}$.

\subsubsection{Even-odd modulation in $xy$-plane}

The first scheme for Floquet engineering serves to suppress the strength of nearest neighbor hoppings relative to next nearest neighbor hoppings in the $xy$-plane.
The modulation takes the form of the even-odd modulation previously mentioned, with $s_x = s_y = 1, s_z = 0$. Specifically, we take
\begin{equation}
\begin{split}
\mu^{xy}_{\b{v}}(t) & = \frac{1}{2} (-1)^{v_x + v_y} \Omega^{xy} g^{xy}  \cos(\Omega^{xy} \, t) \\
\mu^{xy}_{SL}(t) & = \Omega^{xy} g^{xy}_{SL}  \cos(\Omega^{xy} \, t), \\
\end{split}
\end{equation}
where frequency $\Omega^{xy}$ is $2\pi$ times the inverse period, and  $g^{xy}$ and $g^{xy}_{SL}$ are parameters to be tuned. Performing the integral inside Eq.~(\ref{beta AB}) and using $\frac{1}{T} \int_{0}^{T} dt \, \cos \big[ g \sin( 2\pi t/ T) \big] = \frac{1}{2\pi} \int_{0}^{2\pi} dx \, \cos \big[ g \sin(x) \big] = J_0(g)$ gives damping coefficients
\begin{equation}\label{beta xy}
\begin{split}
\beta^{xy,AA}_{\b{r}} & =  
\begin{cases}
J_0(g^{xy})\,\,\,\,\,\,\,\,\,\,\,\,\,\,\,\,\, & r_x + r_y =  \text{odd} \\
1 & r_x + r_y =  \text{even}  \\
\end{cases} \\
\beta^{xy,AB}_{\b{r}} & =  
\begin{cases}
J_0(g^{xy}+g^{xy}_{SL}) & r_x + r_y =  \text{odd} \\
J_0(g^{xy}_{SL})  & r_x + r_y =  \text{even},  \\
\end{cases} \\
\end{split}
\end{equation}
where $J_0(g)$ is a Bessel function of the first kind. We see that `odd' distance hoppings  (including nearest neighbor, $r_x + r_y = 1$) are reduced relative to `even' hoppings (including next-nearest [$r_x = r_y = 1$] and next-next-nearest neighbor [$r_x = 2, r_y = 0$, and vice versa.] hoppings, both with $r_x + r_y=2$). The parameters $g^{xy}$ and $g^{xy}_{SL}$ give independent control over the ratio of even to odd hoppings for both inter- and intra-sublattice hoppings.

\FigureTransitionTwo

\subsubsection{Truncation in the $z$-direction}

The second use of Floquet modulation is to truncate hoppings from power law to short ranged in the $z$-direction~\cite{lee_floquet_2016}.
Unlike the previous $xy$-modulation, we do not have an intuitive explanation for why one needs such a truncation.
Nevertheless, we observe numerically that it is necessary for realizing the Hopf insulator phase. 
We take $\mu^{z}_{\b{v}}(t)$ to be a gradient in the $z$-direction,
\begin{equation}
\begin{split}
\mu^{z}_{\b{v}}(t) & = v_z \Omega^z g^z(\Omega^z t) \\
\mu^{z}_{SL}(t) & = \Omega^z g^z_{SL}(\Omega^z t) \\
\end{split}
\end{equation}
with frequency $\Omega^z$ in time. This gives dampings
\begin{equation}\label{beta z}
\begin{split}
\beta^{z,AA}_{\b{r}} & = \frac{1}{2\pi} \int_{0}^{2\pi} dx \, \cos \bigg[ r_z  \, \int_0^{x} dx' \, g^z(x') \bigg] \\
\beta^{z,AB}_{\b{r}} & = \frac{1}{2\pi} \int_{0}^{2\pi} dx \, \cos \bigg[   \, \int_0^{x} dx' \, [ r_z g^z(x') - g^z_{SL}(x')] \bigg]. \\
\end{split}
\end{equation}
These can be evaluated numerically once the functions $g^z(\Omega_z t), g^z_{SL}(\Omega_z t)$ are chosen. Ref.~\cite{lee_floquet_2016} showed that the modulation can be tuned to give hoppings that are effectively nearest-neighbor in the $z$-direction, at the cost of some loss of magnitude of the nearest-neighbor ($| r_z | = 1$) hopping. For experimental simplicity, we take the modulations to be piecewise constant in time:
\begin{equation} \label{step function}
\begin{split}
\frac{\mu^z_{A,\b{v}}(\Omega_z t)}{\Omega_{z}} & = 
\begin{cases} 
      (2g_1+2) v_z & 0 < \Omega_zt \leq \phi_1/2 \\
      (2g_2+2) v_z & \phi_1< \Omega_zt \leq \phi_2/2 \\
      (2g_3+2) v_z & \phi_2 < \Omega_zt \leq \pi/2 \\
      - \mu^z_{A,\b{v}}(\pi - \Omega_zt) + 4 v_z  & \pi/2 < \Omega_zt \leq \pi \\
      - \mu^z_{A,\b{v}}(2\pi - \Omega_zt) & \pi < \Omega_zt \leq 2\pi \\
\end{cases} \\
\frac{\mu^z_{B,\b{v}}(\Omega_z t)}{\Omega_{z}} & = 
\begin{cases} 
      2 g_1 v_z + 2 g^{SL}_1   & 0 < \Omega_zt \leq \phi_1/2 \\
      2 g_2 v_z + 2 g^{SL}_2 & \phi_1< \Omega_zt \leq \phi_2/2 \\
      2 g_3 v_z + 2 g^{SL}_3 & \phi_2 < \Omega_zt \leq \pi/2 \\
      - \mu^z_{B,\b{v}}(\pi - \Omega_zt)  & \pi/2 < \Omega_zt \leq \pi \\
      - \mu^z_{B,\b{v}}(2\pi - \Omega_zt) \,\,\,\,\,\,\,\,\,\,\,\,\, & \pi < \Omega_zt \leq 2\pi. \
\end{cases}
\end{split}
\end{equation}
Note that $\mu^z_{A,\b{v}}(\Omega_z t)$ is even about $\pi$, guaranteeing that damping coefficients are real-valued [see Eq.~(\ref{beta gen 2})].
The parameters $g_i, g^{SL}_i$ can be tuned to achieve the desired hopping truncation.

\subsubsection{Combining the two modulations}
We now show that both of the above schemes can be implemented simultaneously, by choosing the frequencies of each to be well-separated.
Specifically, we take the modulation to be the sum of two components,
\begin{equation}\label{f full}
\mu^{\alpha}_{\b{r}}(t) = \mu^{xy,\alpha}_{\b{r}}(t) + \mu^{z,\alpha}_{\b{r}}(t),
\end{equation}
where $\mu^{xy,\alpha}_{\b{r}}(t)$ is periodic with frequency $\Omega^{xy}$ and $\mu^{z,\alpha}_{\b{r}}(t)$ with frequency $\Omega^{z}$, and the frequencies satisfy either $\Omega^{xy} \gg \Omega^{z}$ or $\Omega^{xy} \ll \Omega^{z}$. Under this assumption, the dampings $\beta^{\alpha \beta}_{\b{v}}$ factorize into a product of the two individual damping coefficients defined in Eqs.~(\ref{beta xy}) and ~(\ref{beta z}),
\begin{equation}\label{beta full}
\beta^{\alpha \beta}_{\b{v}} = \beta^{xy,\alpha \beta}_{\b{v}} \, \beta^{z,\alpha \beta}_{\b{v}},
\end{equation}
as desired.
We verify that this assumption holds quantitatively in Fig.~\ref{fig: edges}.

\section{Numerical verification of the Hopf insulating phase}\label{sec: numerics}

\FigureTransitionOne

We now turn to a numerical exploration of the single particle bandstructures supported in our dipolar Floquet system.
By tuning the geometric and Floquet engineering parameters, we find transitions from topologically trivial bandstructures to the Hopf insulator and identify parameter regimes where the Hopf insulator's band gap can be as large as $E_g \gtrsim 0.26 t_{\text{nn}}$ (see Figs.~\ref{fig: transition 1},~\ref{fig: transition 2}).  
This occurs with a spacing $a = 0.99$ between adjacent planes of the same sublattice in the $z$-direction (in units of the nearest-neighbor spacing in the $xy$-plane), a spacing $b = 0.66$ between adjacent planes of the opposite sublattice, a staggered chemical potential $\mu^A - \mu^B = 2.28$ (in units of the nearest-neighbor hopping in the $xy$-plane), and Floquet engineering parameters $g^{xy} = 1.2, \, g^{xy}_{SL} = 0.1 , \, g_1 = -0.6, \, g_2 = 0.1,  \, g_3 = 1.1,  \, g^{SL}_1 = 0.7,  \, g^{SL}_2 = -0.4,  \, g^{SL}_3 = 1.6,   \, \phi_1 = 0.2,  \, \phi_2 = 1.8$. 
These optimal parameters were found to optimize the Hopf insulating band gap via the basin-hopping optimization algorithm, a stochastic optimization algorithm that works well in rugged, high-dimensional optimization landscapes.~\cite{wales1997global,2020SciPyNMeth}.
It consists of alternating steps of $i$) performing local optimization to find a nearby local minima in the nearby energy landscape (i.e. the `basin'), and $ii$) randomized `hopping'  to more distant basins, whose local minima are then computed by repeating the first step.
The Floquet engineering amplitudes are quite robust and can be varied together (replacing $g \rightarrow \lambda g$ for all amplitudes defined above) by $\sim \! 25 \%$ about their optimal values while preserving Hopf insulating behavior (Fig.~\ref{fig: transition 1}). The staggered chemical potential can be varied by $\sim \! 20\%$~\cite{schuster2019realizing}.
Performing similar calculations for the lattice parameters, we find that the intra-sublattice distance is also relatively robust and can be varied between $0.5-0.9$ (Fig.~\ref{fig: transition 2}), while the $z$-lattice spacing is slightly more sensitive, and should be kept within $0.92-1.08$ in units of the $x/y$-lattice spacing (note that the most natural value, $1$, lies well-within this range).

We compute the momentum-space Bloch Hamiltonian by summing the Floquet engineered dipolar hoppings defined in Eqs.~(\ref{tSL},~\ref{beta AB},~\ref{beta z}). 
To truncate the infinite sum over hopping distance, we only including hoppings to sites at most $R$ unit cells away in each direction, i.e.~$|r_\mu| \leq R$ for each $ \mu = \{x,y,z\}$.  
The Hopf invariant is computed by evaluating the integral Eq.~(\ref{invariant formula}) on an $N \times N \times N$ grid in momentum space, solving $\nabla \times \b{A}(\b{k}) = \b{j}(\b{k})$ in the inverse Fourier domain to obtain the Berry connection~\cite{moore_topological_2008}. The computed invariant converges quickly to 1 as the discretization $N$ becomes large, e.g. $h - 1 \approx 10^{-6}$ at $N = 70, R = 4$ (see also Figs.~\ref{fig: transition 2},~\ref{fig: transition 1}).
We also see quick convergence of the band gap when increasing $R$, observing quantitative agreement within $10\%$ for all $4 \leq R \leq 32$ and within $1\%$ for all $8 \leq R \leq 32$.

\section{Edge modes of the dipolar Floquet Hopf insulator}\label{sec: edges}

In addition to its linking number invariant, the Hopf insulator's edge modes represent one of its key signatures, and crucially, one which can be experimentally probed.
Up to now, these edge modes are only expected to appear at boundaries that are smooth at the scale of the lattice length, which act as a continuous variation of the two-band momentum-space Hamiltonian $H(\b{k})$ across the boundary region. 
In this case, the Hopf insulator's nontrivial homotopy classification \emph{requires} a gap closing in any edge between the Hopf insulator and the trivial insulator. 
Nevertheless, gapless edge modes have been observed numerically for `sharp' boundaries (i.e.~open boundary conditions)~\cite{moore_topological_2008} and moreover, for the $(001)$-edge, were even shown to be robust to certain perturbations~\cite{deng_hopf_2013}. 

Meanwhile, recent work~\cite{liu_symmetry_2017} has shown that the Hopf insulator's classification can be stabilized to higher bands by a certain crystalline symmetry,
\begin{equation}\label{Cenke}
J H(\b{k})^* J^{-1}  = -H(\b{k}),
\end{equation}
where $JJ^* = -1$, although its classification is reduced to a $\mathbb{Z}_2$ invariant for band number greater than 2. 
This symmetry is in fact automatically satisfied in translationally-invariant two band systems (taking $J = \sigma_y$), and can generally be viewed as the composition of inversion and particle-hole symmetries.

Interestingly, we observe that -- despite involving inversion symmetry -- this crystalline symmetry is also obeyed at the \emph{edge} of a two-band system, in the specific case of a sharp boundary (open boundary conditions).
To see this, note that open boundary conditions are equivalent to an infinite delta function potential barrier at the edge of the system, $H_{\text{edge}} =  \rho \sigma_z \delta_z$, $\rho \rightarrow \infty$, where $\sigma_z$ acts on the sublattice degrees of freedom. 
In momentum space, this potential induces real all-to-all couplings between different values of $k_z$, $H^{\b{k},\b{k}'}_{\text{edge}} = \rho \sigma_z \delta_{k_x,k'_x} \delta_{k_y,k'_y}$.
This is now easily seen to obey Eq.~(\ref{Cenke}) with $J = \sigma_y$.

\FigureEdges

This observation suggests that the edge modes previously observed at sharp boundaries of the Hopf insulator are in fact protected by this `unintentional' crystalline symmetry, and are therefore not robust to perturbations that break the symmetry.
To test this, we solve for the (100)-edge modes of the dipolar Hopf insulator via exact diagonalization for three different edge terminations: sharp, sharp with a symmetry-breaking perturbation, and adiabatic. We observe three qualitatively distinct spectra [Fig.~\ref{fig: edges}(a-c)]. 
The sharp edge hosts a linear energy degeneracy, consistent with previous studies~\cite{moore_topological_2008,deng_hopf_2013}.
To break the crystalline symmetry, we add a site-dependent chemical potential $\mu_{\b{v}} \mathbbm{1}$ localized on the two unit cells $\b{v}$ nearest the edge. 
This perturbation gaps the edge mode, supporting our conjecture that the sharp edge modes of the Hopf insulator are in fact crystalline-symmetry-protected~\cite{fn8}.

Finally, we consider smooth boundaries between the Hopf insulator and the trivial insulator. 
To construct smooth boundaries, we take the hoppings to be constant throughout the lattice, while an $x$-dependent staggered chemical potential $\mu_x \sigma_z$ tunes the Hamiltonian between the trivial phase at each end of the lattice and the Hopf insulating phase in the center. 
This interpolation occurs smoothly over two `edge regions' on either side of the Hopf insulating phase, consisting of $\sim 20$ lattice sites each. 
Shown in Fig.~\ref{fig: edges}(c), these smooth edges also feature gapless edge modes.
Importantly, the gaplessness of these edge modes is robust to \emph{any} smooth perturbation to the lattice, including a `smoothed' version of the site-dependent chemical potential that was observed to gap the sharp edge mode [Fig.~\ref{fig: edges}(d)].

\section{Experimental Proposal}\label{sec: expt}

We now turn to our central result: a detailed blueprint for realizing the dipolar Hopf insulator using ultracold polar molecules. 
An explosion of recent experimental progress has led to the development of numerous possible molecular species \cite{ni2008high,park2015ultracold,takekoshi2014ultracold,guo2016creation,molony2014creation}, but for concreteness (and to demonstrate that the requisite separation of energy scales can be quantitatively realized), here we focus on ${}^{40}$K$^{87}$Rb~\cite{ni2008high,moses2015creation,yan_observation_2013,ospelkaus_controlling_2010,aldegunde_manipulating_2009,aldegunde_hyperfine_2008}. 

We begin with the geometry and rotational level diagram illustrated in Fig.~\ref{fig: lattice}. 
The 3D optical lattice is generated using four pairs of counter propagating beams, two forming the $xy$-lattice and two forming the $A$ and $B$ sublattices in the $z$-direction.
For experimental convenience, we envision the two sublattices to be formed by beams with orthogonal linear polarizations of light. 
In this case a birefringent mirror can control the relative phase between the two reflected beams, which in turn determines the separation between sublattices.

To realize the rotational level diagrams of Fig.~\ref{fig: lattice}, we first propose to tune the rotational states $\ket{1,0}$ and $\ket{1,1}$ of all molecules to be approximately degenerate using applied DC electric and magnetic fields, oriented in the $z$-direction with amplitudes 1650 V/m and -490 G respectively~\cite{fn3}. 
The degeneracy between the $\ket{1,0}$ and $\ket{1,1}$ states, and, in turn, the sublattice symmetry between the $A$ and $B$ planes, can then be broken by using \emph{different} intensities of light to form each sublattice.
Owing to the AC polarizability of ${}^{40}$K$^{87}$Rb, the lattice beams not only trap the molecules in the designated geometry, but also induce $m_J$-dependent shifts in the molecules' rotational states proportional to the beams' intensities~\cite{neyenhuis_anisotropic_2012}. 
The individual intensities, $I_A$ and $I_B$, can therefore be tuned such that the transitions $\ket{1,0}_A \leftrightarrow \ket{0,0}_A$ and $\ket{1,1}_B \leftrightarrow \ket{0,0}_B$ are near-resonant with each other, yet off-resonant with all other transitions.
Specifically, we calculate that $x$-polarized light with intensities $I_A = $0.43 kW/cm$^2$ and $I_B = $0.54 kW/cm$^2$ leads to the desired near-resonance, with an energy gap $\delta \sim 5$ kHz to the nearest rotational state outside the prescribed doublets.
Energy levels are calculated as in Ref.~\cite{neyenhuis_anisotropic_2012}, and we assume the $x$- and $y$-lattices are formed with $z$-polarized light of intensity .5 kW/cm$^2$.
The molecule ${}^{40}$K$^{87}$Rb has a rotational splitting $\Delta = 2.2$ GHz and a measured dipolar interaction strength $t \sim 50$ Hz when trapped in a 3D optical lattice with $1064 $nm light~\cite{yan_observation_2013}. 
This scheme therefore naturally leads to the desired separation of energy scales $t \ll \delta \ll \Delta$.

\FigureExperimentFloquet

Energy levels in hand, let us turn to the implementation of the Floquet modulations (Fig.~\ref{fig: experiment}). 
To realize the $xy$-plane modulation, we can again rely upon the AC polarizability, using a two-dimensional intensity-modulated standing wave to directly tune the molecules' energy levels non-uniformly in both space and time.
The energy shifts of the $\ket{1,0}_A$ and $\ket{1,1}_B$ states can be made equal [necessary to ensure the modulation is of the form of Eq.~(\ref{Floquet mu form})] by tuning the polar angle of the light's polarization to $\theta = .96$ rad, owing to the anisotropic polarizability of ${}^{40}$K$^{87}$Rb~\cite{neyenhuis_anisotropic_2012}. 
An additional stationary standing wave on the even sites can cancel the site-dependent non-zero average of the modulation, preserving translation invariance of the effective chemical potential. 
At a modulation frequency, $\Omega_{xy} \sim 500$ Hz, much greater than the dipolar interaction strength, $t_{\text{nn}} \sim 50$ Hz, the optimal modulation strength $g^{xy} = 1.2$ requires an intensity $\sim 10^{-2}$ kW/cm$^2$. 
An additional space-independent modulation of the two beams enables a difference between the two sublattices' modulations, achieving a nonzero $g^{xy}_{SL}$.

This method does not work for the $z$-gradient Floquet modulation, as a $z$-gradient in the light's intensity is necessarily accompanied by a polarization in the orthogonal $xy$-plane. 
In addition to shifting the molecules' energy levels, such a polarization would also induce mixing between rotational states, contaminating the desired hopping phase structure.
Rather, we propose to achieve the $z$-gradient Floquet modulation by combining two independent sources of modulation [Fig.~\ref{fig: polarizability}(c-e)]. 
First, we apply an oscillating electric field gradient of order $\delta E / \delta z \sim  1$ kV/cm$^2$. 
This gradient alone is not sufficient to realize the modulation of Eq.~(\ref{Floquet mu form}), because it shifts the energies of the the $\ket{1,0}_A$ and $\ket{1,1}_B$ states differently, owing to their different polarizability.
We therefore combine this with a circularly-polarized beam tuned \emph{near}, but off-resonant with, the $^3\Pi_{0^+}$ electronic excited state of ${}^{40}$K$^{87}$Rb, which shifts the energy levels of the low-lying rotation states of interest via the AC Stark shift [Fig.~\ref{fig: polarizability}(a-c)].
We imagine the beam to be traveling in the $z$-direction, with the natural transverse spreading of the beam along its propagation axis giving rise to a $z$-gradient in intensity $\delta I(z)/\delta z \sim I(z)/z$~\cite{yariv1991optical}.
To this end, we perform calculations of the AC polarizabilities of ${}^{40}$K$^{87}$Rb with circularly-polarized light as a function of detuning from the $b^3\Pi_{0^+}$ state [Fig.~\ref{fig: polarizability}(b)] using experimentally adjusted potential energy curves~\cite{Alps2016,Pashov2007} as well as parallel and perpendicular electronic polarizabilities~\cite{neyenhuis_anisotropic_2012}, which we expand on in detail in the following section.
For $\sigma^+$ light, the polarizabilities have poles at the resonant transition frequency to the excited $J=2$ state, which allows the corresponding energy 
shifts to be precisely controlled by the detuning over a large range.
Modulating the detuning about resonance (as a step function, to avoid any resonance-induced decay) precisely realizes the desired Floquet modulation. 
Quantitatively, we find that detunings $\Delta \nu \sim 1$ GHz lead to AC polarizibilities $\alpha/h \sim 1$ kHz/(W/cm$^2$), which in turn requires intensity gradients $\delta I / \delta z \sim $ 5  W/($\mu$m cm$^2$) to achieve the optimal Floquet parameters at modulation frequency $\Omega_z \sim 5 \text{ kHz } \gg \Omega_{xy}$.
At a distance $z \sim 100 \, \mu$m, the desired intensity gradient is thereby achieved with a modest intensity $I \sim .5$ kW/cm$^2$ and power $P \sim I(z)\times z^2 \sim 50$ mW~\cite{yariv1991optical}.

We do not expect our proposed Floquet modulations to introduce substantial heating to the molecular system for a number of reasons. First, the modulations occur at a frequency significantly faster than the Hamiltonian energy scales, which exponentially suppresses many-body energy absorption~\cite{abanin2015exponentially}. Second, since the Hopf insulator's topology is characterized via its single-particle band structure, one only needs to excite a small number of molecules at any given time. At this single-particle level, the primary concern turns to heating from parametric processes associated with the laser intensity modulation. 
In this case, one can again utilize a separation of energy scales, by choosing the frequencies of the Floquet modulation to be far removed from any trap resonances (i.e. the trap frequency and its harmonics) such that no parametric heating will take place~\cite{neyenhuis_anisotropic_2012,2012PhDT89N}.
Typical values of the trap frequency for ${}^{40}$K$^{87}$Rb experiments are $\sim \! 20$ kHz with a quality factor $\sim \! 20$~\cite{2012PhDT89N}; resonances are therefore easily avoided both in our simple order-of-magnitude estimate, $\Omega_z \sim 5$ kHz and  $\Omega_{xy} \sim 500$ Hz, as well as our more quantitative estimate in Fig.~\ref{fig: edges}, using $\Omega_z = 600 t_{\text{nn}} \approx 30$ kHz and  $\Omega_{xy} = 25 t_{\text{nn}} \approx  1.25$ kHz.

The edge modes of the dipolar Hopf insulator can be probed experimentally via molecular gas microscopy~\cite{marti2018imaging,covey2018approach}.
Here, a tightly-focused beam applied near the edge induces local differences in the molecules' rotational splittings, enabling one to spectroscopically address and excite individual dipolar spins. 
The extent to which such an excitation remains localized on the edge during subsequent dynamics can be read out using spin-resolved molecular gas microscopy.
For polar molecules separated by a distance of $1 \, \mu$m, single-molecule addressing of the $\ket{0,0} \rightarrow \ket{1,0}$ transition has been estimated to require a beam of radius $1 \,\mu$m and a reasonable power $10\,\mu$W~\cite{covey2018approach}.
The width of the edge region, typically large due to a wide harmonic confining potential, can be tuned via a number of recently developed techniques, including: box potentials~\cite{gaunt2013bose}, additional `wall' potentials~\cite{zupancic2016ultra}, or optical tweezers~\cite{liu2018building}, allowing one to realize the three scenarios depicted in Fig.~\ref{fig: edges}. 

\FigurePolarizability

\section{Details on AC polarizabilities for \lowercase{$z$}-direction modulation}\label{app: circular}

To effectively implement the Floquet modulation along $z$-direction,
we use circularly polarized light tuned near a narrow transition, which
allows light shifts to be precisely controlled by the detuning from
the transition. Specifying to the molecule ${}^{40}$K$^{87}$Rb, we choose the
dipole-forbidden transition 
$|X^1\Sigma^+,v=0,J=1,m_J\rangle \to |b^3\Pi_{0^+},v=0,J=2,m_J+1\rangle$
with $1028.7$ nm~\cite{Kobayashi2014} $\sigma^+$ light 
where $m_J=0$ for the A sublattice and $1$ for the B sublattice.
With relatively weak laser intensity (on the order of W/cm$^2$), the 
light shift can be characterized by the AC polarizability of the
molecular state of interest. The polarizability is calculated from 
two different contributions. 
The first and more important contribution comes from the
resonant transition which has a strong dependence on the detuning, and
the second contribution comes from all other transitions that has 
negligible dependence on the detuning in the range we are interested in.
Here we assume the detuning is much larger
than the spacings between $|X^1\Sigma^+,v=0,J=1,m_J\rangle$ states with
$m_J = 0$ and $\pm 1$, and these spacings are much larger than the
light shifts.

To characterize the contributions from the resonant transition, we 
follow the recipe in Refs.~\cite{Kotochigova2006,KBonin1997,AJStone1996}.
The generally complex dynamic polarizability for alkali-metal molecule in 
a rovibrational state of the ground $X^1\Sigma^+$ potential is given by
\begin{equation}\label{eq:alpha}
\begin{split}
\alpha(h & \nu,\hat{\varepsilon}) = \\
& \frac{1}{\varepsilon_0 c}
 \sum_f \frac{E_f - E_i - ih\gamma_f/2}
             {(E_f - E_i - ih\gamma_f/2)^2-(h\nu)^2}
 |\langle f|d\hat{R}\cdot\hat{\varepsilon}|i\rangle|^2
\end{split}
\end{equation}
$\hat{\varepsilon}$ and $\nu$ are the polarization vector and
the frequency of the light, respectively, $c$ is the speed of light, 
$\varepsilon_0$ is the electric constant, $\hat{R}$ is the 
orientation of the interatomic axis, and $d$ is the dipole operator.
$i$ denotes the rovibrational state $|i\rangle$ of interest with energy
$E_i$ in the ground $X^1\Sigma^+$ potential, and the summation over 
$f$ denotes the summation over all rovibrational states $|f\rangle$ 
other than $i$ with energies $E_f$ in all electronic potentials, and 
$\gamma_f$ describe the natural linewidths of $|f\rangle$.

When the laser frequency is very close to the narrow dipole-forbidden
transition, the most significant contribution comes from that transition
which has a pole at the resonant frequency and weakens as the inverse 
function of the detuning.
We treat all transitions from $|X^1\Sigma^+,v=0,J=1,m_J\rangle$
to rovibrational states in the $b^3\Pi_{0^+}$ potential
using Eq.~(\ref{eq:alpha}). The largest contribution by far
comes from the transition to the excited $v=0$ state due to the
similarity of its radial wavefunction to the ones in the ground
potential. We use the experimentally adjusted
potential energy curves for both the excited $b^3\Pi_{0^+}$ 
state~\cite{Alps2016} and the ground $X^1\Sigma^+$ 
state~\cite{Pashov2007}, and a spin-orbit modified 
transition dipole moment between them~\cite{Kotochigova2004}.
Since the natural linewidths of the lowest rovibrational states in
the $b^3\Pi_{0^+}$ potential are much smaller (on the order of 
kHz~\cite{Kobayashi2014}) then the detunings we are interested in
(on the order of GHz), we take $\gamma_f = 0$.

The background contributions from all other transitions have 
negligible frequency dependence close to the
$1028.7$ nm transition due to the large detunings from the 
corresponding excited states.
Thus we treat the background polarizabilities as constants throughout
the detuning range. We use the method in Ref.~\cite{Kotochigova2010}
with experimentally determined electronic parallel and perpendicular
polarizabilities~\cite{neyenhuis_anisotropic_2012} to calculate the
background polarizabilities at $1064$ nm and assume them to be the
same near the $1028.7$ nm transition. More specifically, we use 
$\alpha_\parallel/h=10.0(3)\times 10^{-5}$ MHz/(W/cm$^2$) and 
$\alpha_\perp/h=3.3(1)\times 10^{-5}$ MHz/(W/cm$^2$) determined for 
the wavelength of $1064$ nm and obtain the background polarizabilities
$\alpha_{bg,|1,0\rangle}/h=4.64\times 10^{-5}$ MHz/(W/cm$^2$) and 
$\alpha_{bg,|1,1\rangle}/h=5.98\times 10^{-5}$ MHz/(W/cm$^2$) for
$\sigma^+$ polarization. 

Finally, we add the two parts together to arrive at the total AC
polarizabilities shown in Fig.~\ref{fig: polarizability}(b).

\section{Conclusions}

We have completed our specification of how Hopf insulating phases can be realized and detected in near-term experiments on ultracold polar molecules.
As one of the few known topological insulators to fall outside both the traditional tenfold way classification as well as its extension to crystalline symmetries, the Hopf insulator is a particularly interesting phase of matter with many open questions eager for experimental input.
For instance, we have proposed using the presence of a gapless edge mode at a smooth boundary, probed by spectroscopy, as a robust experimental diagnostic of the Hopf insulating phase.
Recent work suggests that at the $(001)$-edge this mode should feature a nonzero Chern number associated with an unusual bulk-to-boundary flow of Berry curvature~\cite{alexandradinata2019actually}; numerous techniques to measure the Chern number have been developed~\cite{price2012mapping,aidelsburger2015measuring,wimmer2017experimental,tarnowski2019measuring}, which may allow one to detect this physics.
Looking to the future, an experimental Hopf insulator would be a vital resource in the search for a bulk response characterized by the Hopf invariant (analogous to the Hall effect in a Chern insulator), which so far remains unknown.

Our blueprint may also provide a basis from which to realize various extensions of the Hopf insulator.
In our proposal, we have already seen that polar molecules can realize certain crystalline symmetry-protected extensions of the Hopf insulator~\cite{liu_symmetry_2017,alexandradinata2019actually}, which can be detected independently from the ordinary (non-crystalline) Hopf insulator by looking at sharp edge terminations that respect the crystalline symmetry.
Polar molecules might also be used to realize driven extensions of the Hopf insulator, for instance, the Floquet Hopf insulator~\cite{schuster2019floquet}.
Here, one subjects the system to periodic driving at a time-scale \emph{comparable} to the hopping time, which can lead to a new Floquet Hopf insulating phase, characterized by a $\mathbbm{Z} \times \mathbbm{Z}_2$ pair of topological invariants that underlie an even richer spectrum of edge mode behavior than in the non-driven case.
The Floquet Hopf insulator can be realized by strobing a flat band static Hopf insulator with periodic $\pi/2$-pulses of a staggered chemical potential~\cite{schuster2019floquet} -- the latter would be easily realized via a $\sim\!100$ Hz oscillation of the lattice light intensity.
Realizing a sufficiently flat band Hopf insulator is a less trivial task, but the bandwidth could be optimized via standard optimization techniques depending on the specific set of available experimental parameters.
More speculatively, a flat band Hopf insulator might also be a natural launching ground into many-body generalizations of the Hopf phase (much as a flat band Chern insulator is a key ingredient for the fractional Chern insulator~\cite{bergholtz2013topological}), which are thus far unexplored territory.

In the context of polar molecules, our work applies a number of tools developed for controlling and cooling polar molecules towards quantum simulation.
We hope that some selection of these tools may find broader utility.
For instance, our use of a sublattice-dependent lattice light intensity to realize (pseudo)spin-orbit coupling via the $\Delta m = 1$ component of the dipolar interaction may prove fruitful in realizing other topological phases as well.
As a simple example to demonstrate wider applicability, the exact same form of spin-orbit coupling ($t^{AB}_{\textbf{r}} \sim e^{i\phi}$) in 2D gives rise to Chern insulating physics~\cite{qi2006topological}.
In polar molecule setups limited by the ability to fill only a (random) fraction of the full set of lattice sites, the Chern insulator might therefore provide a disorder-robust~\cite{agarwala2019topological} stepping stone to realizing the Hopf insulator.
We have also provided implementations of two independent Floquet engineering schemes: an even-odd patterning utilizing the molecules' AC polarizability under lattice light, and a truncation of the power-law dipolar interaction in the $z$-direction via a single circularly-polarized Gaussian laser beam.
Floquet engineering has proven critical in other quantum simulation platforms, and these techniques may serve as building blocks for its use in polar molecules. 
At a higher level, our work provides yet another piece of evidence for the power of dipolar interaction, and the potential of polar molecules as a quantum simulation platform. 

\emph{Acknowledgments}---We gratefully acknowledge the insights of and discussions with Dong-Ling Deng, Luming Duan, Vincent Liu, Kang-Kuen Ni, and Ashvin Vishwanath. This work was supported by the AFOSR MURI program (FA9550-21-1-0069), the DARPA DRINQS program (Grant No. D18AC00033), NIST, the David and Lucile Packard foundation, the W. M. Keck foundation, and the Alfred P. Sloan foundation. T.S. acknowledges support from the National Science Foundation Graduate Research Fellowship Program under Grant No. DGE 1752814. F. F. acknowledges support from a Lindemann Trust Fellowship of the English Speaking Union, and the Astor Junior Research Fellowship of New College, Oxford. Work at Temple University is supported by ARO Grant No. W911NF-17-1-0563, AFOSR Grant No. FA9550-21-1-0153, and NSF Grant No. 1908634. 

\bibliographystyle{apsrev4-1}
\bibliography{refs_DipolarHopf,refs_SvetlanaMing}

\begin{thebibliography}{97}
\bibitem [{fn2()}]{fn2}%
  \BibitemOpen
  \href@noop {} {\bibinfo  {journal} {In the presence of external fields, which
  slightly mix the rotational eigenstates, we use $\ket{J,m_J}$ to refer to the
  state adiabatically connected to the associated rotational eigenstate}\
  }\BibitemShut {NoStop}%
\end{thebibliography}

\end{document}